\documentclass[12pt]{article}
\usepackage{graphicx}
\usepackage{hyperref}

\newcommand\pubnumber{SNSN-XXX-YY}
\newcommand\pubdate{\today}

\def\nwu{Department of Physics and Astronamy\\
Northwestern University, Evanston, IL, USA}

\def\Title#1{\begin{center} {\Large #1 } \end{center}}
\def\Author#1{\begin{center}{ \sc #1} \end{center}}
\def\Address#1{\begin{center}{ \it #1} \end{center}}

\newcommand\pubblock{\rightline{\begin{tabular}{l} \pubnumber\\
      \pubdate  \end{tabular}}}
\newenvironment{Abstract}{\begin{quotation}  }{\end{quotation}}
\newenvironment{Presented}{\begin{quotation} \begin{center} 
      PRESENTED AT\end{center}\bigskip 
    \begin{center}\begin{large}}{\end{large}\end{center} \end{quotation}}

\begin{document}
\begin{titlepage}
\pubblock

\vfill
\Title{Fermionic decays of SM Higgs }
\vfill
\Author{Andrey Pozdnyakov\\on behalf of the ATLAS and CMS collaborations}
\Address{\nwu}
\vfill
\begin{Abstract}
In this document I present an overview of the recent results published by ATLAS and CMS collaborations
on the searches for SM Higgs boson decay to fermions.
The document summarizes the status of the analyses up to September of 2014
and contains the results of pp collision Data at $\sqrt{s} = $7~and~8\,TeV.
Searches for $H\to\tau\tau$, $H\to bb$, $H\to\mu\mu$ and $ttH$ processes are presented.

\end{Abstract}
\vfill
\begin{Presented}
XXXIV Physics in Collision Symposium \\
Bloomington, Indiana,  September 16--20, 2014
\end{Presented}
\vfill
\end{titlepage}
\def\thefootnote{\fnsymbol{footnote}}
\setcounter{footnote}{0}

\section{Introduction}
\begin{figure}[tb]
  \centering
  \includegraphics[width=0.75\textwidth]{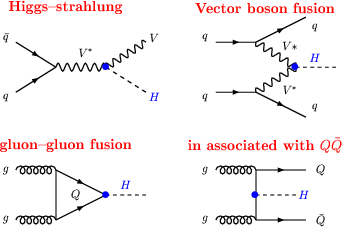}
  \caption{Diagrams of Higgs production processes. Image from \cite{higgs-theory}}
  \label{fig:dia}
\end{figure}

The existence of the scalar boson with mass of 125\,GeV has been established. 
The properties of this new particle thus far are consistent with the standard model (SM) Higgs boson.
Decay modes for its discovery, $H \to ZZ/WW$ and $H \to \gamma\gamma$, provide an indirect evidence for 
Higgs coupling to top quark due to its production in gluon-gluon fusion process (Fig.~\ref{fig:dia}).
Nevertheless a direct evidence for its decay to fermions is crucial in order 
to uncover the true nature of the new particle.  
Standard Model predicts that the coupling of the Higgs to fermions is proportional to the mass of the fermion,
hence one expects larger branching fraction of the decay to heavier leptons.
For example for a 125 GeV Higgs SM predicts $\mathcal{B}(H \to bb) = 58\,\%$, $\mathcal{B}(H \to \tau\tau) = 6\,\%$.

In proton-proton collisions at LHC leading Higgs boson production mechanisms are (Fig.~\ref{fig:dia}):
gluon-gluon fusion (ggF) -- about 88\% for SM Higgs with $m_H=125$ GeV at $\sqrt{S} = 8$\,TeV;
Vector Boson Fusion (VBF) -- 7\%; associated production with a Z or W boson (VH) -- 5\%;
and a $t\bar{t}$ fusion (ttH) -- 0.4\%. 

Even though ggF process dominates, there are experimental advantages of VBF and VH modes: 
tagging events with extra particles and reducing the backgrounds.
In VH production the tag is based on the leptonic decays of the Z/W bosons:
missing transverse energy from neutrinos ($E_T^{miss}$) 
and/or leptons -- W($\ell\nu$)H, Z($\ell\ell$)H and  Z($\nu\nu$)H.
Let me note the branching ratios of those decays: $\mathcal{B}(W\to\ell\nu) \approx 10\%$ per lepton,
$\mathcal{B}(Z\to\ell\ell) \approx 3.4\%$, $\mathcal{B}(Z\to\nu\nu) \approx 20\%$.

Typical VBF tag requires an event with two jets with $m_{jj} > 500$\,GeV
and $|\eta_{j_1}~-~\eta_{j_2}|~>~3.5$.

This Note summarizes recent results released by ATLAS and CMS experiments up to September of 2014.
It includes analyses of the pp Data sets at $\sqrt{s} = 7$\,TeV with $\mathcal{L}_{int} \approx 5 fb^{-1}$
and $\sqrt{s} = 8$\,TeV with $\mathcal{L}_{int} \approx 20 fb^{-1}$.

\section{\texorpdfstring{$H \to \tau\tau$}{Higgs to tau tau}}
The search for $H\to\tau\tau$ decay is challenging experimentally due to several reasons:
(1)~reconstruction of the $E_T^{miss}$ from neutrinos, which is difficult at hadronic colliders;
(2)~jet reconstruction and energy resolution in hadronic final states;
(3)~large irreducible background from $Z\to\tau\tau$ process.

In addition the analysis complicates by three different final states due to decay modes of the taus:
$H\to\tau_{lep}\tau_{lep}$ (12\%), $H\to\tau_{lep}\tau_{had}$ (46\%), $H\to\tau_{had}\tau_{had}$ (42\%).

\begin{figure}[tb]
  \centering
  \includegraphics[width=.45\textwidth]{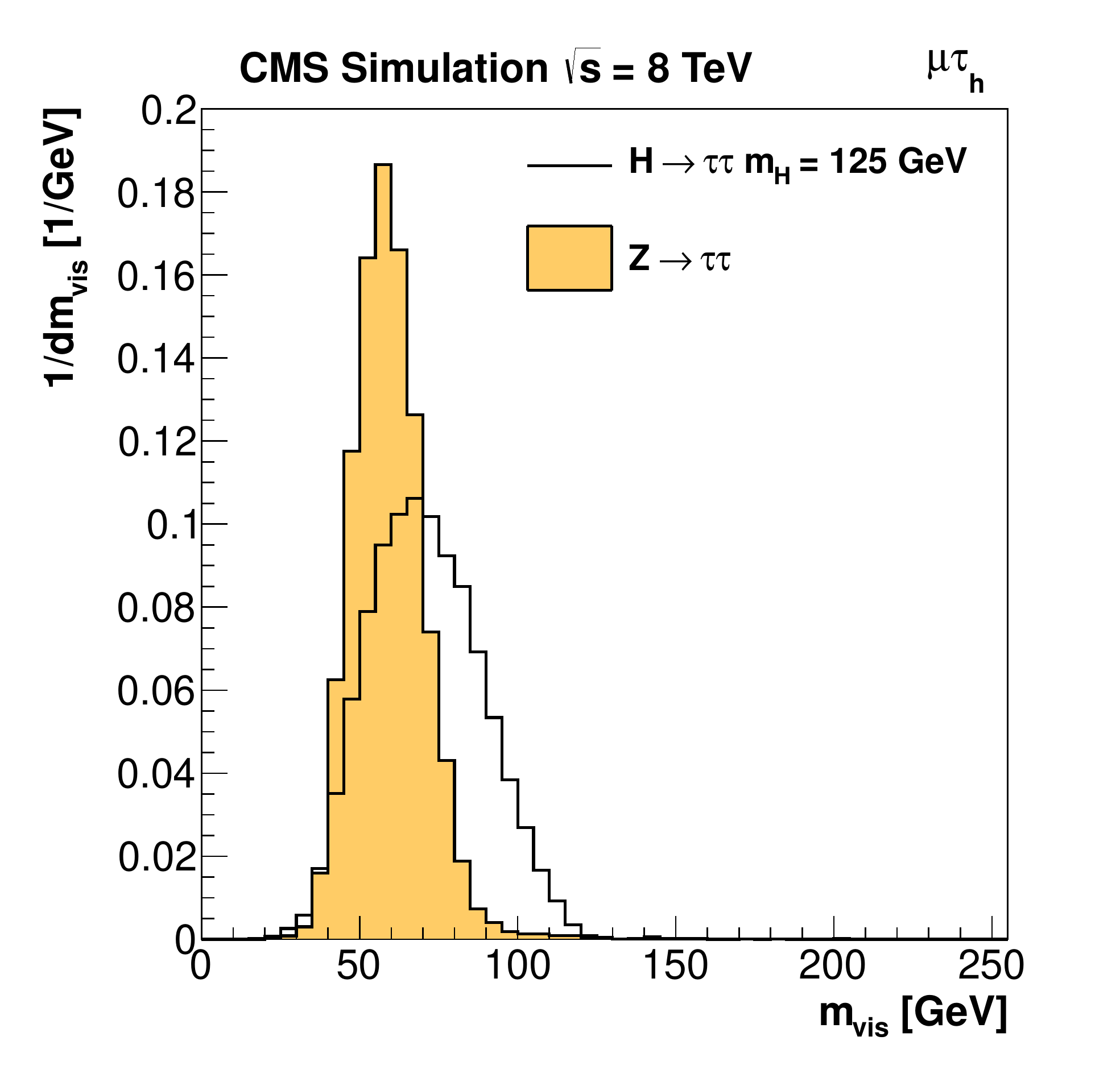}~
  \includegraphics[width=.45\textwidth]{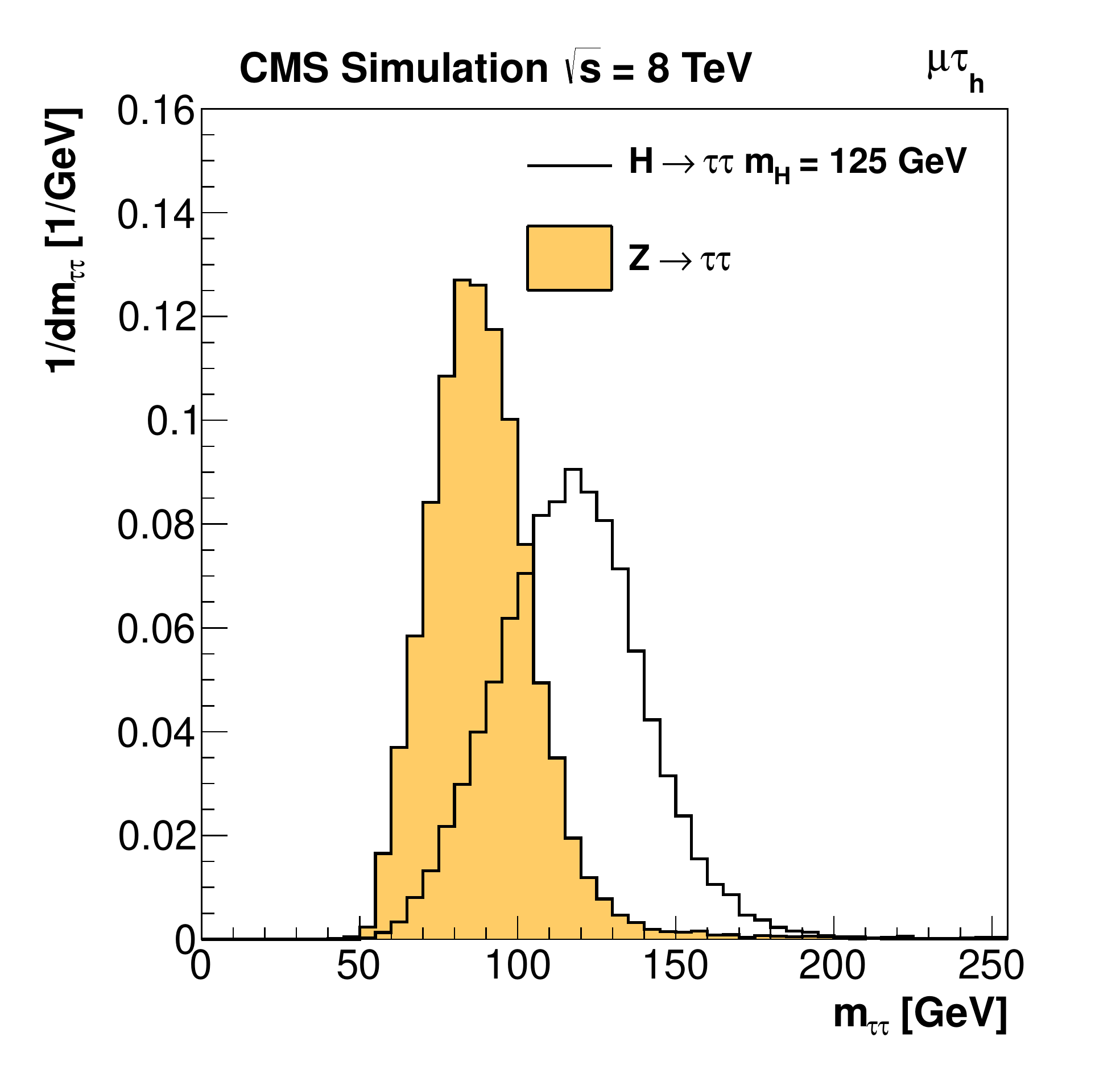}
  \caption{CMS: invariant mass of $\tau\tau$ from $Z\to\tau\tau$ and $H\to\tau\tau$ MC samples,
  reconstructed from ``visible'' particles (left) and after full recovery (right).}
  \label{fig:cms-Htautau-1}
\end{figure}

Reconstruction of the Higgs boson candidate mass, $m_{\tau\tau}$, from the visible decay products 
of the $\tau$-lepton  is one of the key ingredients of the analysis.
Both ATLAS and CMS accomplish this with similar methods: 
ATLAS makes use of Missing Mass Calculator algorithm~\cite{atl-Htautau},
while CMS does matrix element Likelihood Function minimization~\cite{cms-Htautau}.
See Fig.~\ref{fig:cms-Htautau-1} for the validation of this procedure from CMS.

Both ATLAS and CMS estimate the contribution from $Z\to\tau\tau$ background 
using so-called \textit{embedding} technique: 
they take well identified $Z\to\mu\mu$ events in Data and insert the $\tau$-leptons instead of the muons in those events.
Tau particles are then decayed by TAUOLA program (which takes into account polarization) 
and detector simulation of those decays is inserted into the Data events. 
This technique is shown to perform very well with various validation tests both in Data and Monte Carlo (MC).

\subsection{ATLAS Analysis}
In order to increase sensitivity of the analysis events are
separated into mutually independent categories based on the presence of jets and their topology (VBF),
and $p_T^{\tau\tau}$ of the Higgs candidate.
Major backgrounds (besides already mentioned $Z\to\tau\tau$) are estimated from Data 
(different methods used in different final states).

After basic pre-selection of events BDT training is performed with 6-9 most important variables 
(including $m_{\tau\tau}$). Separate BDTs are trained for each final state and event category.
Then, a global likelihood fitting is done using the BDT scores and event rates in the control regions.

\subsection{CMS Analysis}
Additional categories are introduced (compared to the ATLAS analysis) based on 
the number of jets, VBF and VH topology, and $p_T^{\tau\tau}$.
Shape and yield of major backgrounds are estimated from Data:
$Z\to\tau\tau$ from the embedding method; QCD from same-sign lepton events.
For some other backgrounds the shape is taken from MC while normalization 
is from Data: $Z\to\mu\mu/ee$, W+jets, $t\bar{t}$.

A global fit is performed using $m_{tt}$ variable for most categories (while BDT is used
for $H\to\tau_{\mu}\tau_{\mu}$ $H\to\tau_e\tau_e$  final states).

\subsection{Results for \texorpdfstring{$H \to \tau\tau$}{Higgs to tau tau}}
Once the final background composition is determined from the global fit in control regions
one looks into the signal region to determine the amount of excess (if any).
Both experiments do see an excess of events over the backgrounds with significance greater than $3\sigma$
and the signal strength of $1.4\pm0.4$ (ATLAS) and $0.8\pm0.3$ (CMS),
which constitutes the evidence for $H\to\tau\tau$ decay! 

\begin{figure}[t]
  \centering
  \includegraphics[width=.45\textwidth]{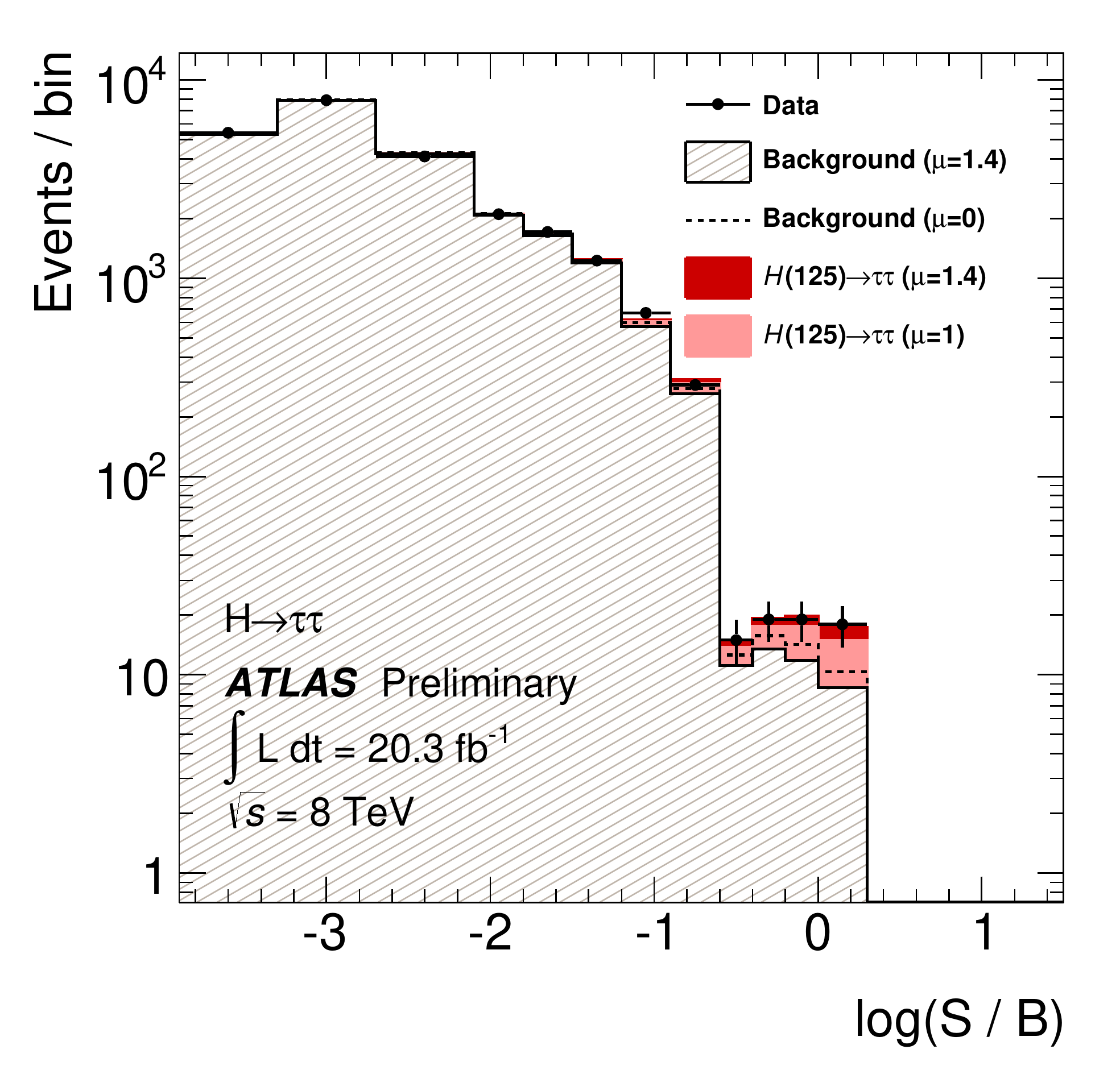}~
  \includegraphics[width=.45\textwidth]{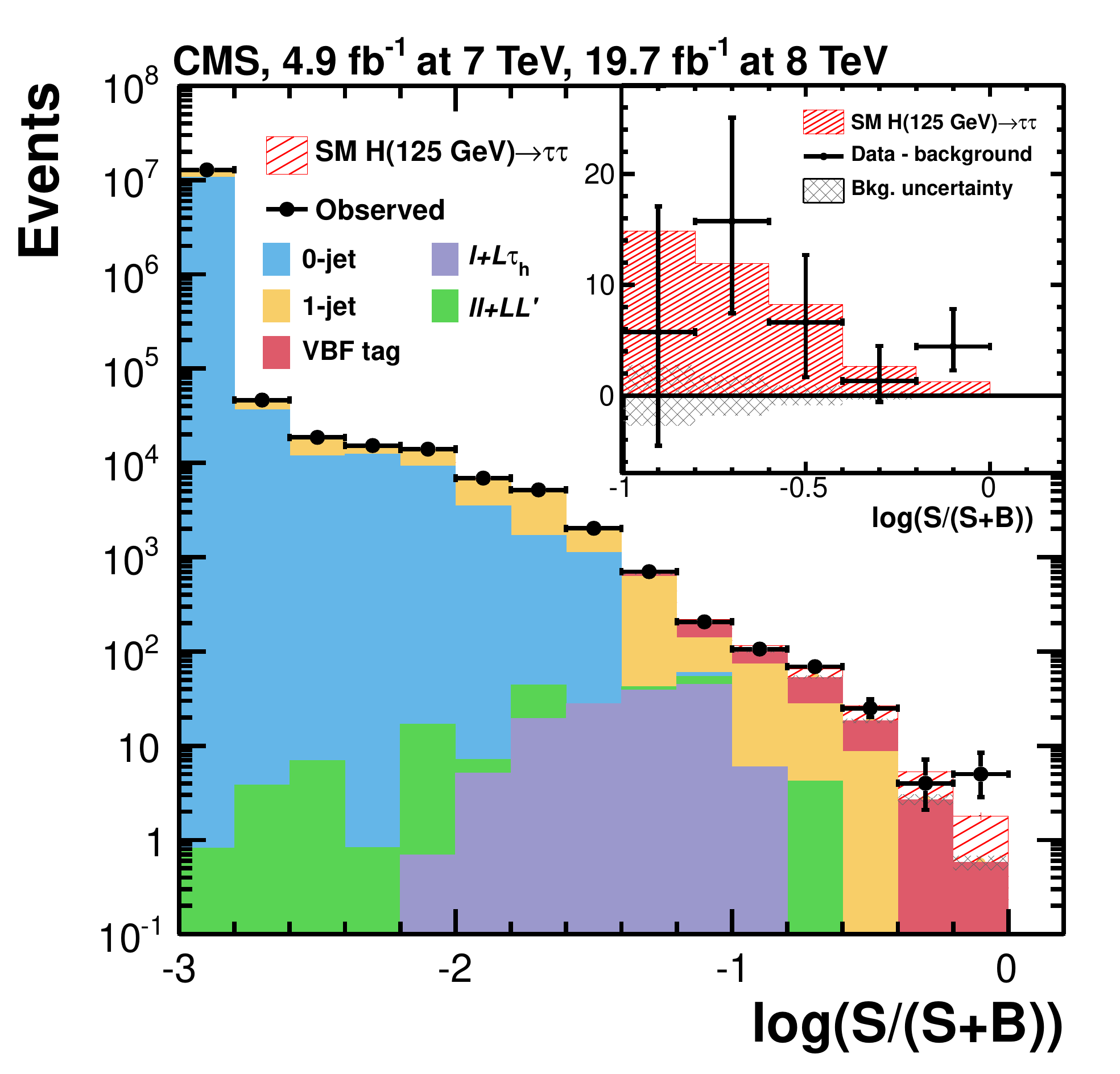}
  \caption{$H\to\tau\tau$ candidate events in bins of expected S/B ratio, ATLAS (left) and CMS (right).}
  \label{fig:Htautau-SB}
\end{figure}

\begin{figure}[t]
  \centering
  \includegraphics[width=.6\textwidth]{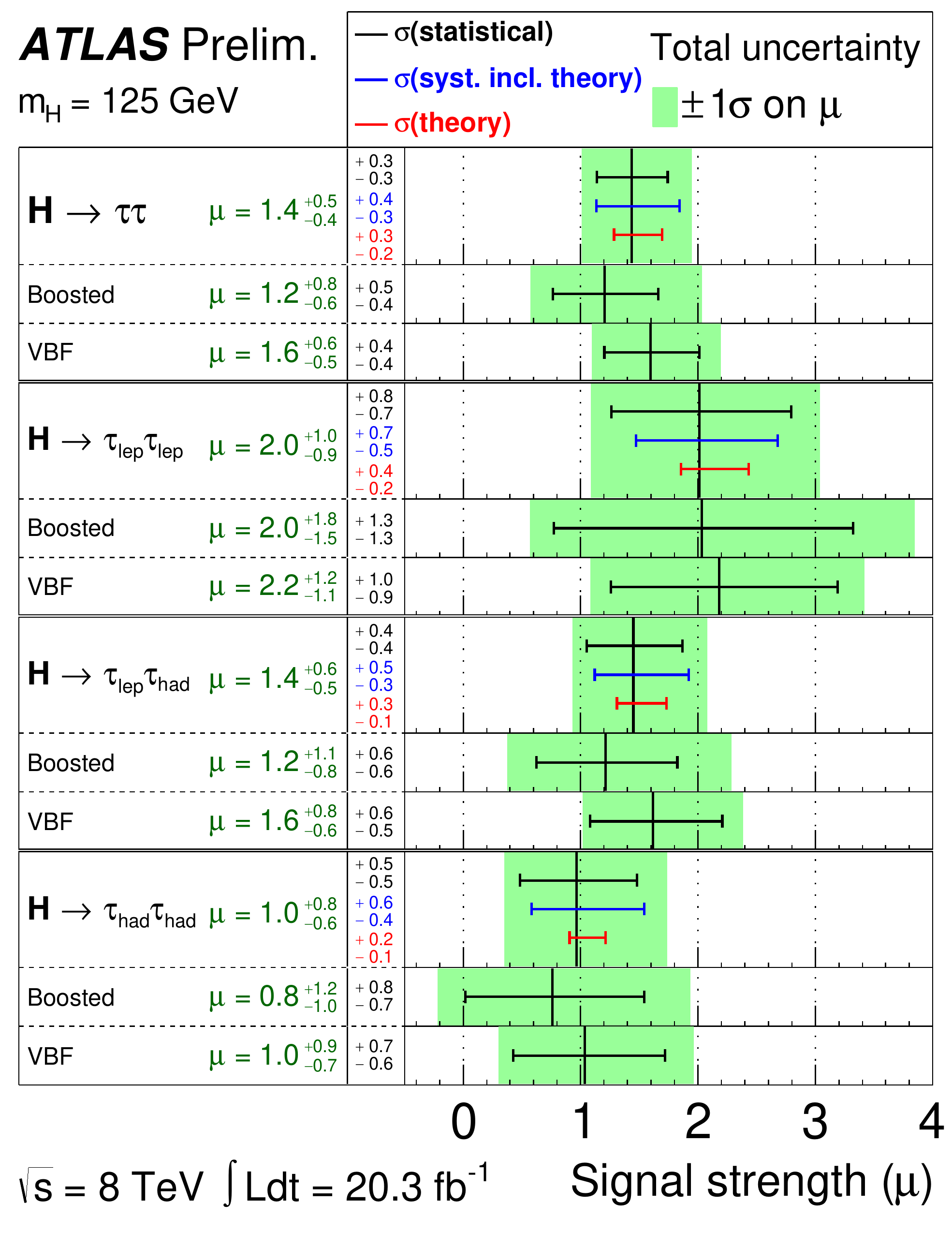}
  \caption{ATLAS $H\to\tau\tau$ results on signal strength.}
  \label{fig:atl-Htautau-res}
\end{figure}

\begin{figure}[b]
  \centering
  \includegraphics[width=.4\textwidth]{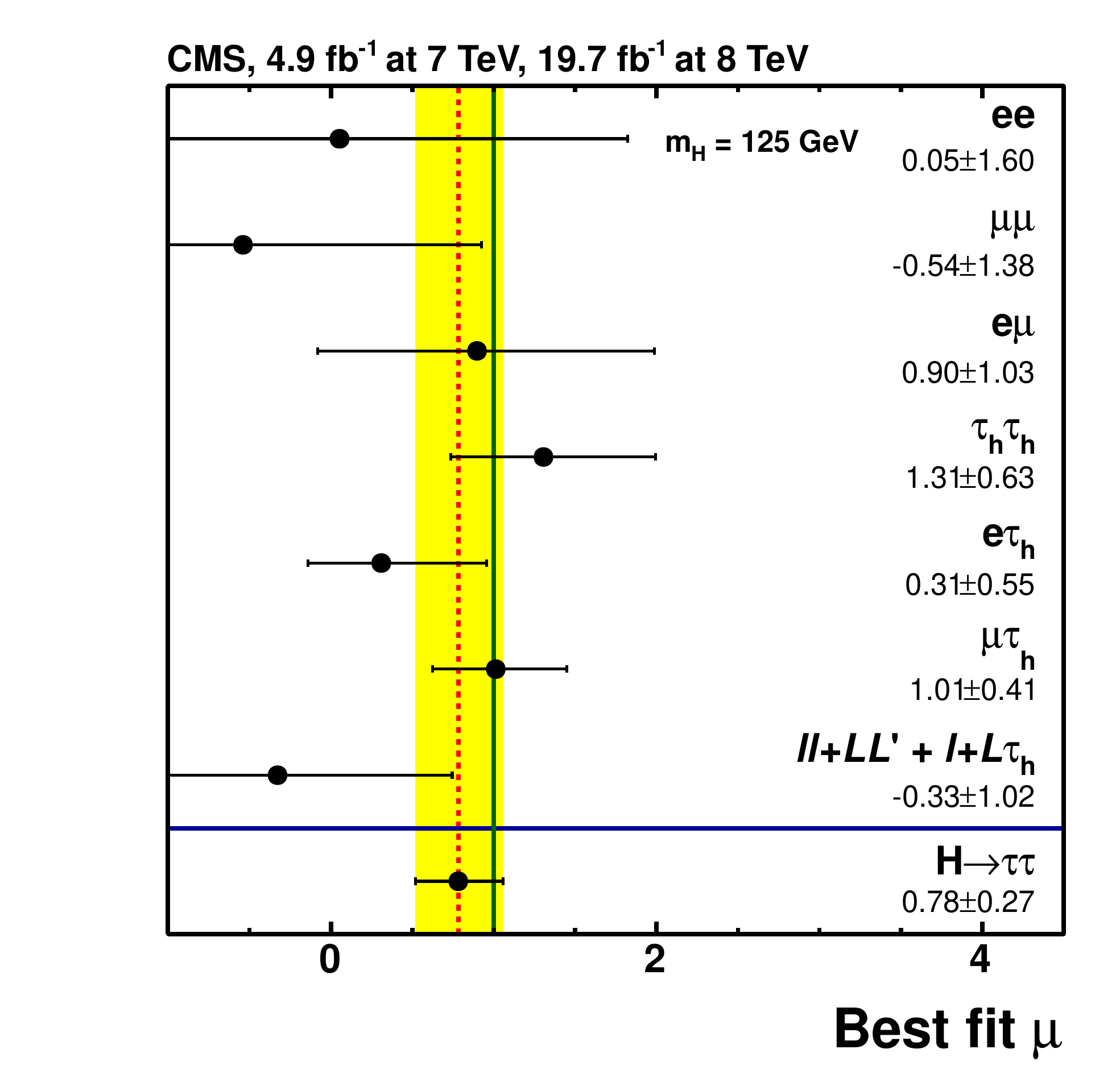}
  \includegraphics[width=.4\textwidth]{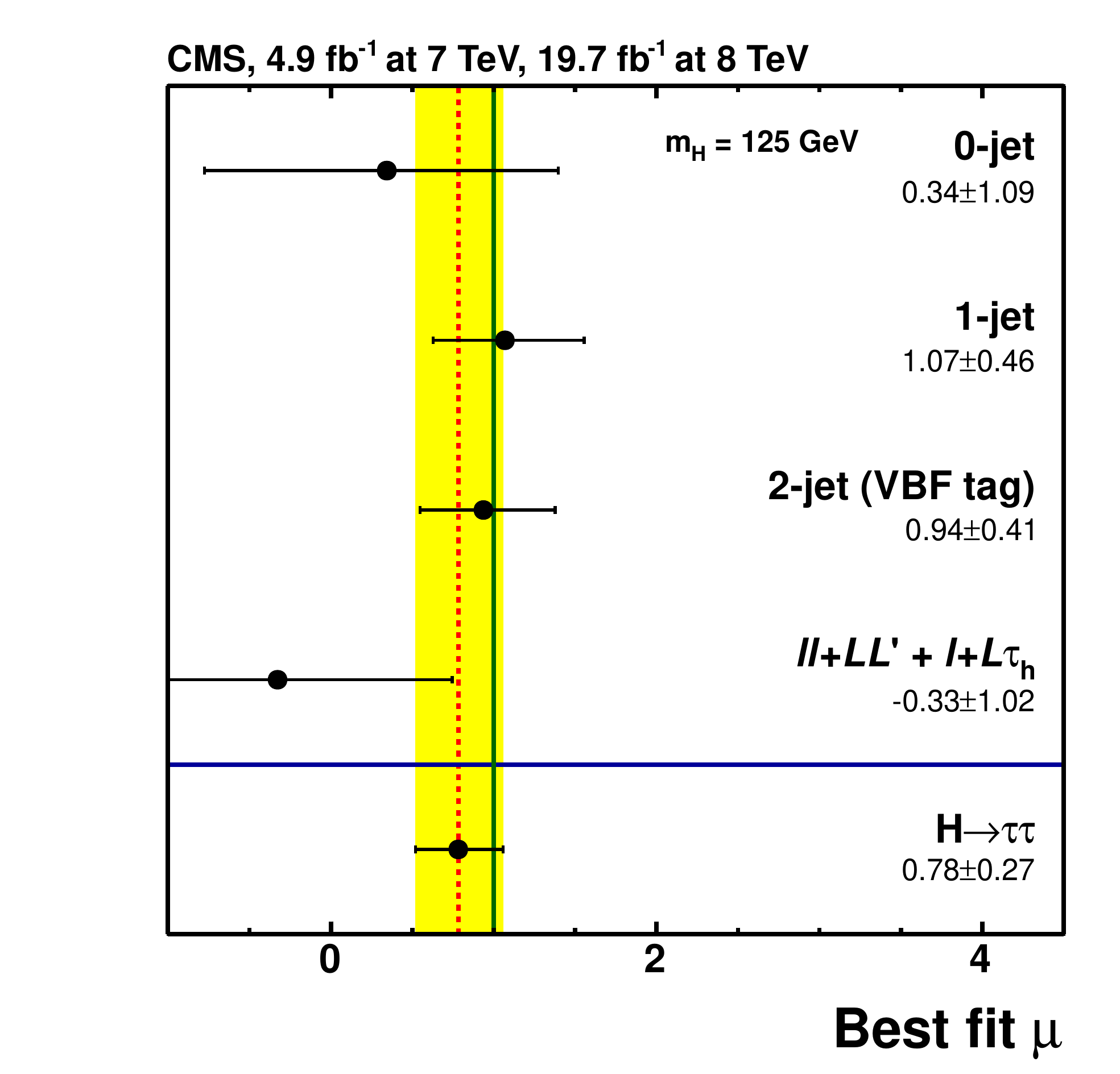}
  \caption{CMS $H\to\tau\tau$ results on signal strength.}
  \label{fig:cms-Htautau-res}
\end{figure}

See Figs.~\ref{fig:atl-Htautau-res} and \ref{fig:cms-Htautau-res} for the details on the signal strength 
across all the analysis categories. The breakdown of uncertainties from ATLAS
is instructive -- systematics dominate and the largest one is the theoretical uncertainty  
(due PDFs, ``scale'', NNLO corrections). 
Another visually attractive way to present the results is the distributions
in Fig.~\ref{fig:Htautau-SB}, where events are binned 
according to their expected Signal-to-Background (S/B) ratio. 

\section{\texorpdfstring{$H \to b\bar{b}$}{Higgs to bb}}
Very large background rate prevents doing this search in ggF channel. 
However, tagging events with an extra lepton and/or $E_T^{miss}$ enables the analysis in VH channel,
which is the way both ATLAS and CMS experiments go \cite{atl-VHbb, cms-VHbb}.
(VBF channel is also possible but has lower sensitivity than VH (CMS); it is not presented here.)
Besides the VH tagging, a key feature of the analysis is the possibility 
to identify the jets from b-quarks due to a displaced vertex. 
Efficiency of the b-jet ID varies but one can think of $\varepsilon\sim70\%$ for simplicity.

Both ATLAS and CMS perform a BDT type of analysis.
As a validation of the methods used in the Higgs boson search they also measure 
SM cross section of $pp\to VZ$ process with $Z \to b\bar{b}$ decay.
It is found to be consistent with SM prediction~\cite{atl-VZbb, cms-VHbb}.

\begin{figure}[t]
  \centering
  \includegraphics[width=.35\textwidth]{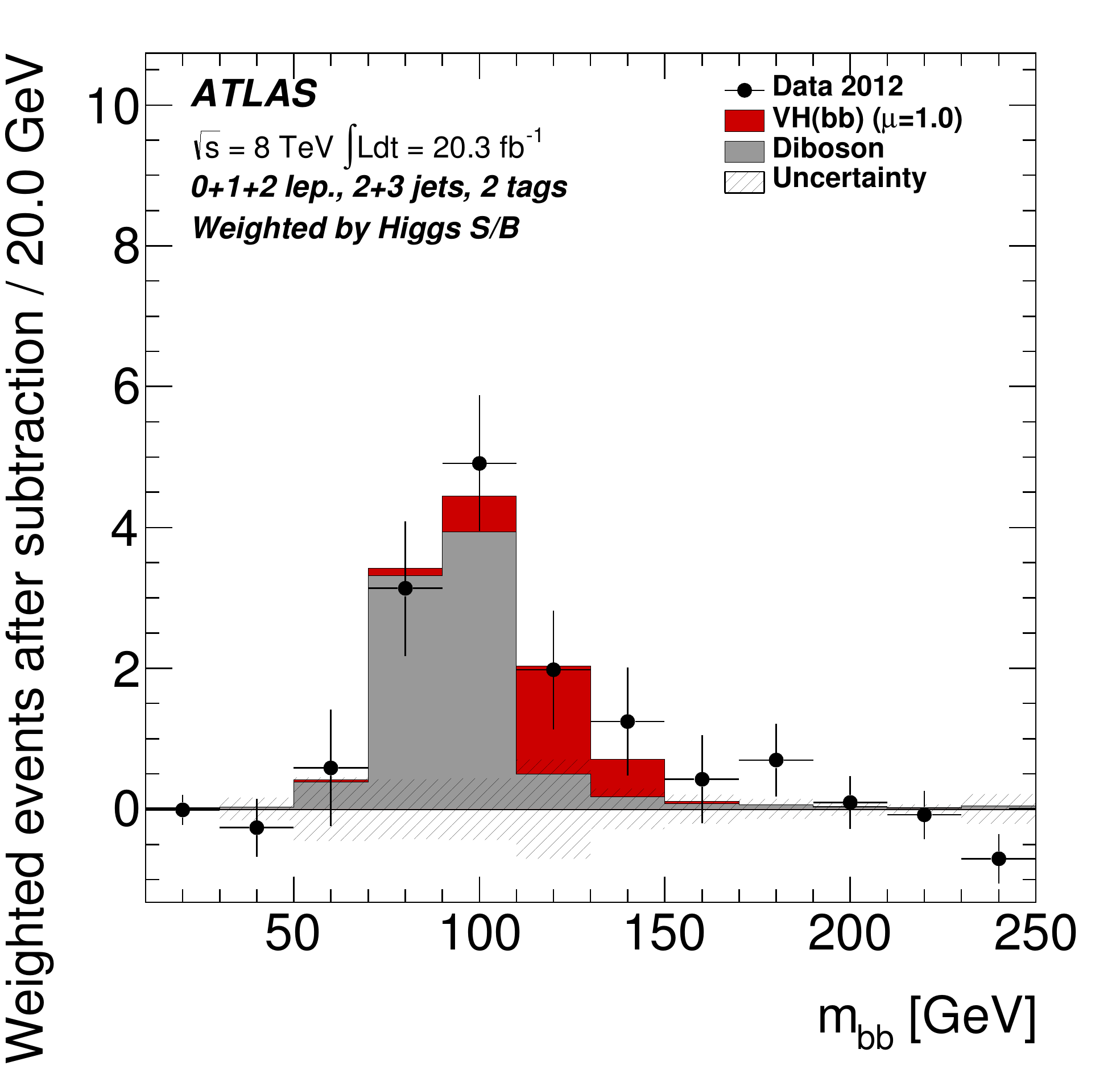}~
  \includegraphics[width=.35\textwidth]{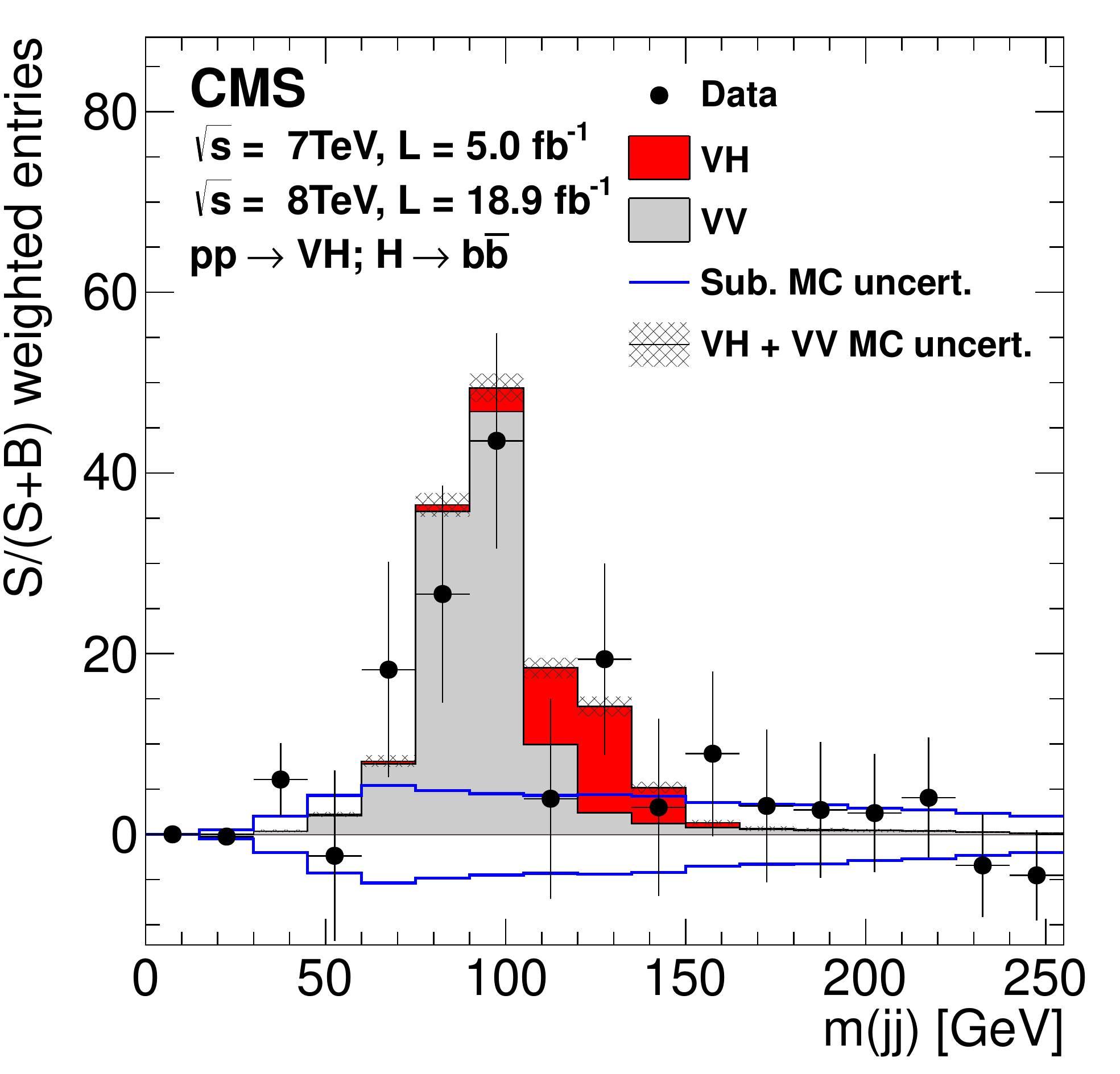}
  \caption{Final distribution of $m_{bb}$ where events 
    are weighted according to their S/B bin (Fig.~\ref{fig:cms-Htautau-res});
    ATLAS (left) and CMS (right).}
  \label{fig:Hbb-mbb}
\end{figure}

\begin{figure}[t]
  \centering
  \includegraphics[width=.45\textwidth]{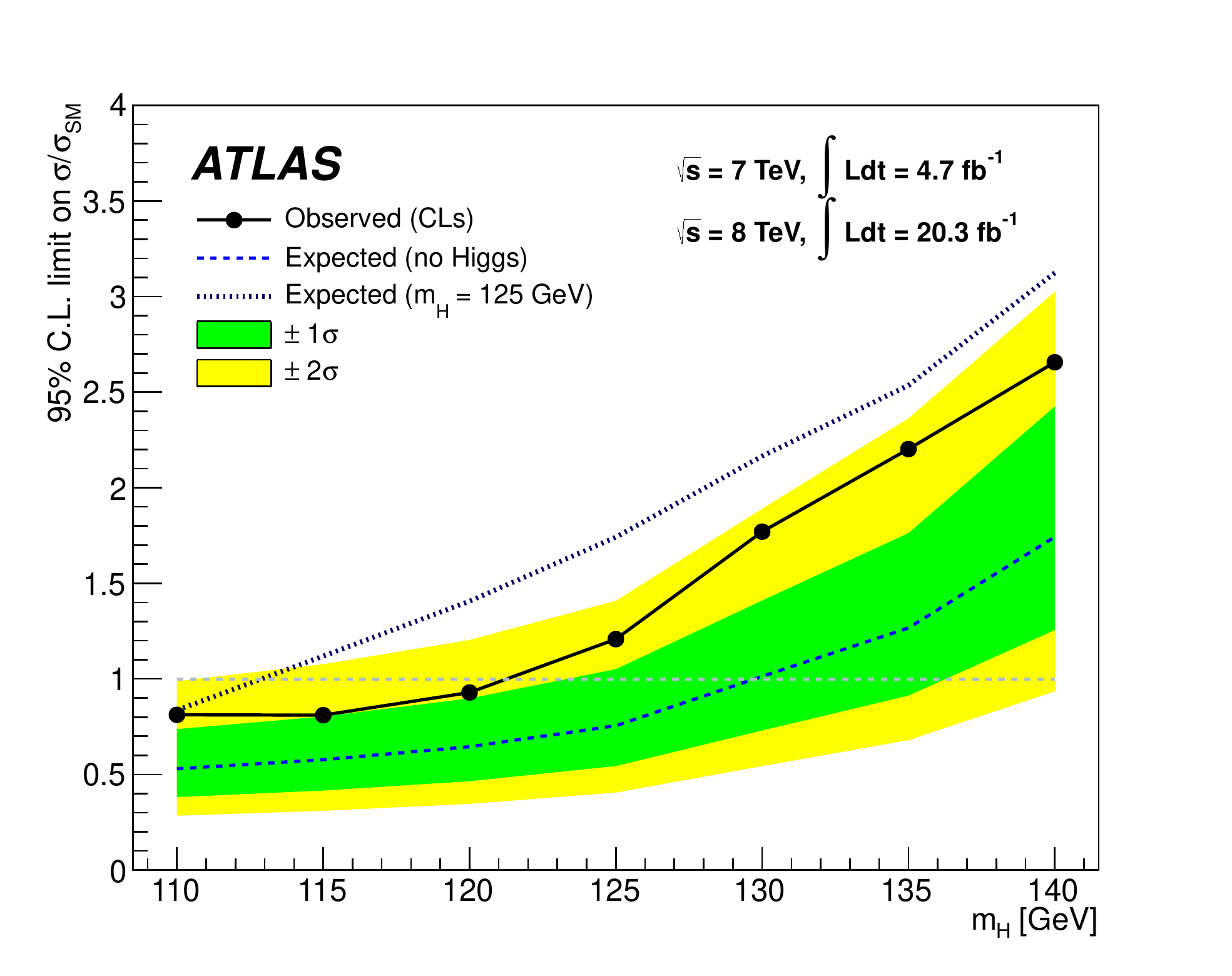}~
  \includegraphics[width=.35\textwidth]{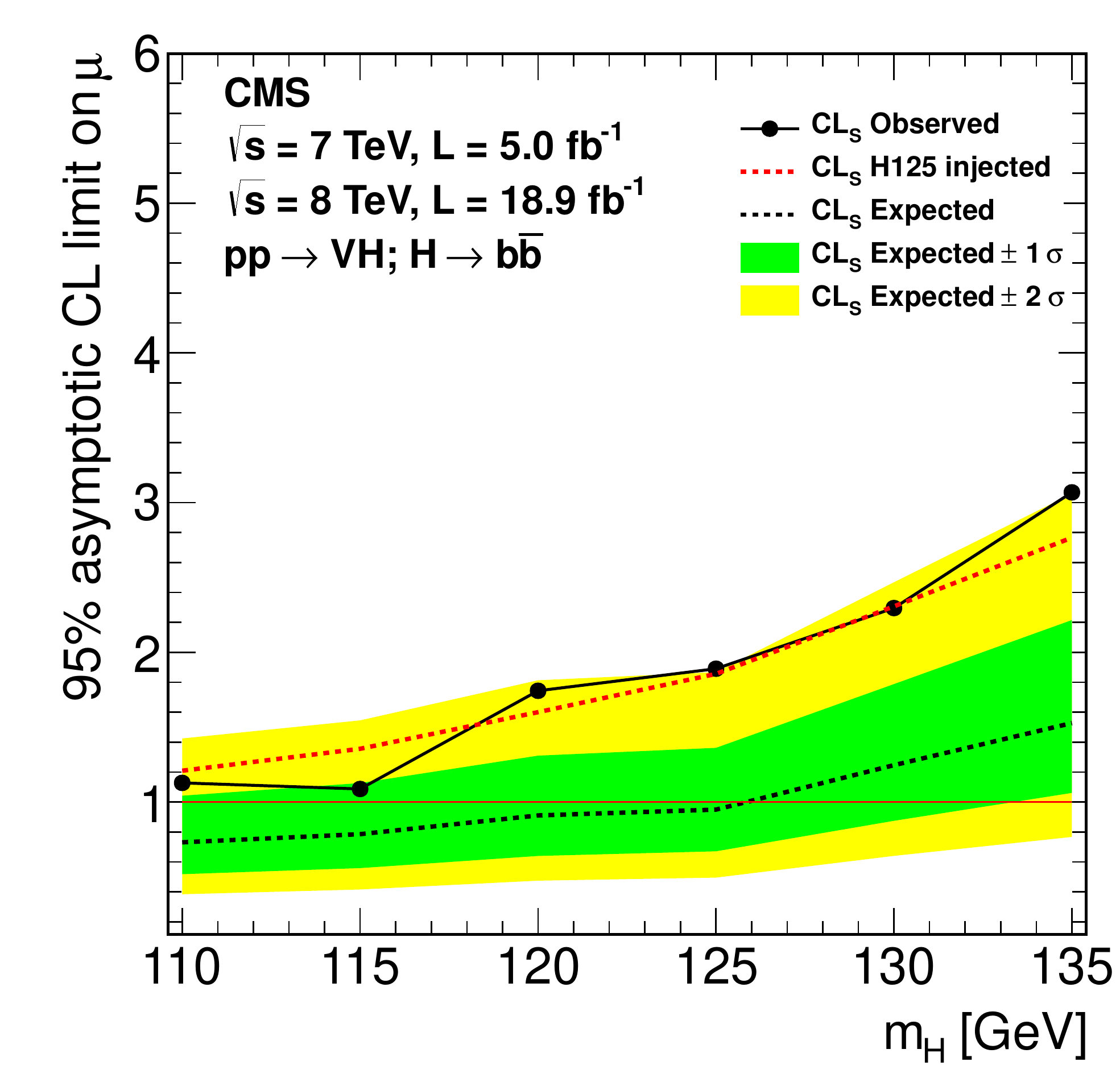}
  \caption{Upper limits on the signal strength of the $VH(bb)$ 
    by ATLAS (left) and CMS (right).}
  \label{fig:Hbb-lim}
\end{figure}

Expected upper limit in both experiments is less than $1\times$SM,
however an excess of events observed in Data is not yet significant to claim the evidence for $H\to bb$ decay. 
Final di-jet mass distributions are shown in Fig.~\ref{fig:Hbb-mbb} 
and the results (upper limits) are in Fig.~\ref{fig:Hbb-lim}.

\section{\texorpdfstring{$H\to\mu\mu$}{Higgs to mu mu}}
Decay of the Higgs to a pair of muons provides a clean final state 
and requires a rather simple analysis: one looks for a bump in the dimuon invariant mass, $m_{\mu\mu}$, spectrum.
It is however a very small signal in SM, $\mathcal{B}(H \to\mu\mu) = 2*10^{-4}$,
and has large irreducible background from $Z \to \mu^+\mu^-$ events.

Both experiments enhance the sensitivity of the search with event categorization~\cite{atl-Hmumu}~\cite{cms-Hmumu}.
Here, for example, is a basic set up by ATLAS: (1) VBF tag, (2) Non-VBF category which is split further into
2 subcategories based on $\eta^{\mu}$ and 3 subcategories based on $p_T^{\mu\mu}$.
(Both muons in $|\eta^{\mu}|<1$ or otherwise; three $p_T^{\mu\mu}$ bins: (0--15), (15--50) and $>50$\,GeV.)

Background is estimated from a smooth fit to Data events.
In non-VBF categories the background model is the sum of a Breit-Wigner (BW) function 
convoluted with a Gaussian (GS), and an exponential function divided by $x^3$:
$\mathcal{P}(x) = f\cdot[BW(m, \Gamma)*GS(\sigma)](x)+(1-f)Ce^{Ax}/x^3$;
while for VBF events it is the product of a Breit-Wigner and an exponential function:
$\mathcal{P}(x) = BW(m,\Gamma, x)\cdot e^{A·x}$, where $x\equiv m_{\mu\mu}$. 

Signal model is obtained from MC via a fit to a Crystal Ball plus Gaussian function with the same mean value. 

\begin{figure}[b]
  \centering
  \includegraphics[width=.35\textwidth]{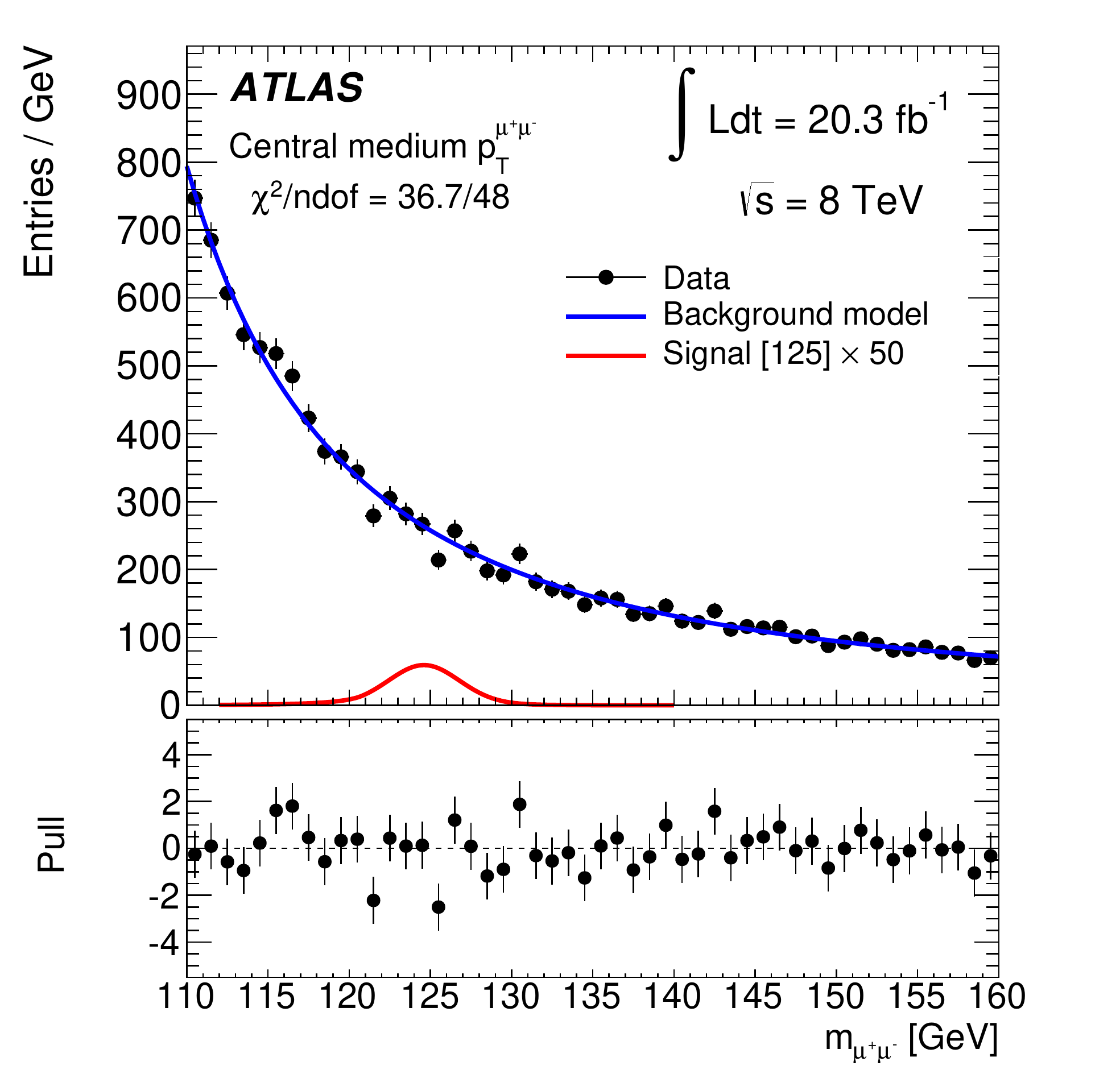}~
  \includegraphics[width=.35\textwidth]{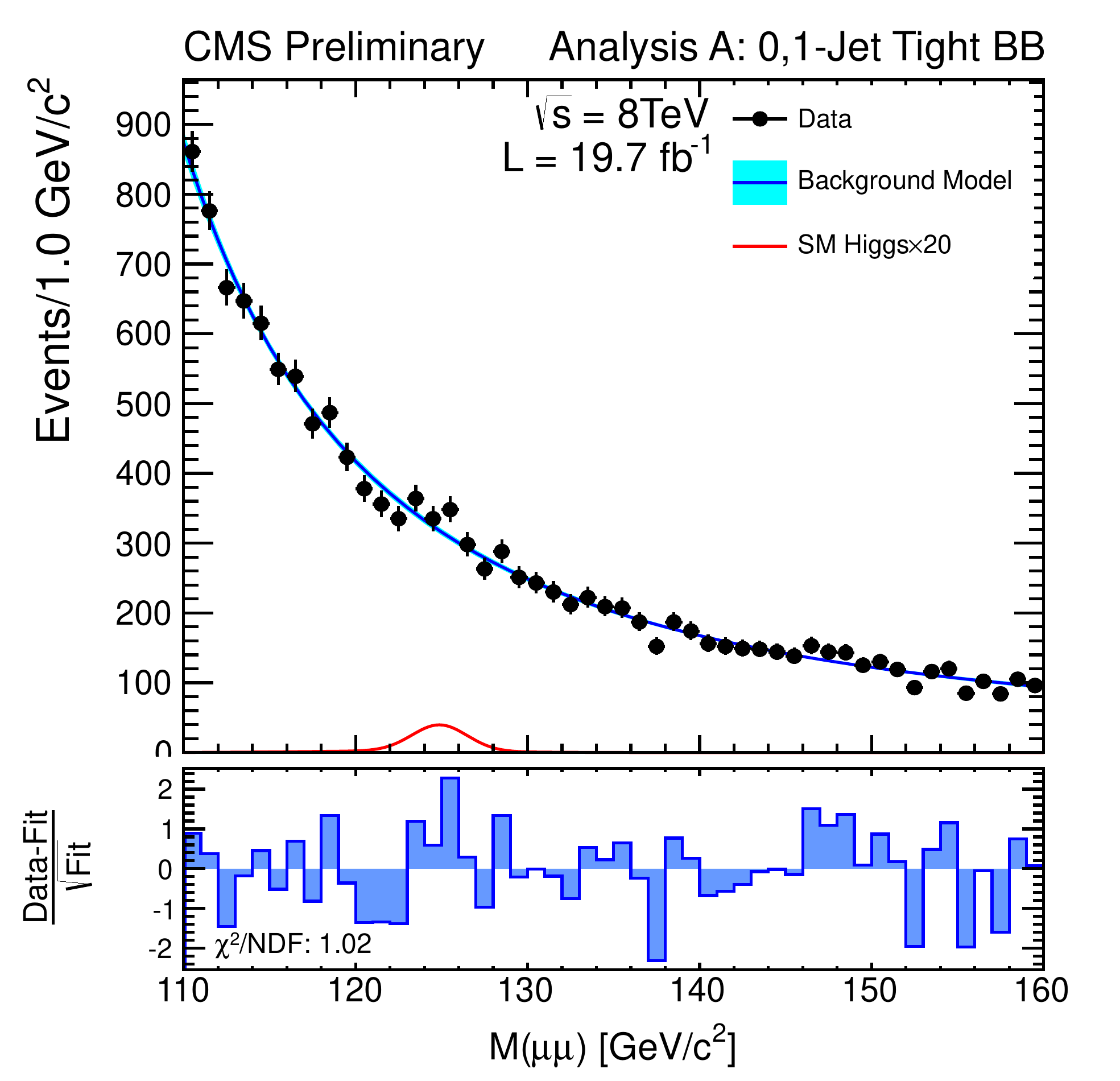}
  \caption{$m_{\mu\mu}$ distributions of  ``central'' events by ATLAS (left) and  CMS (right).}
  \label{fig:Hmumu-mass}
\end{figure}

\begin{figure}[t]
  \centering
  \includegraphics[width=.55\textwidth]{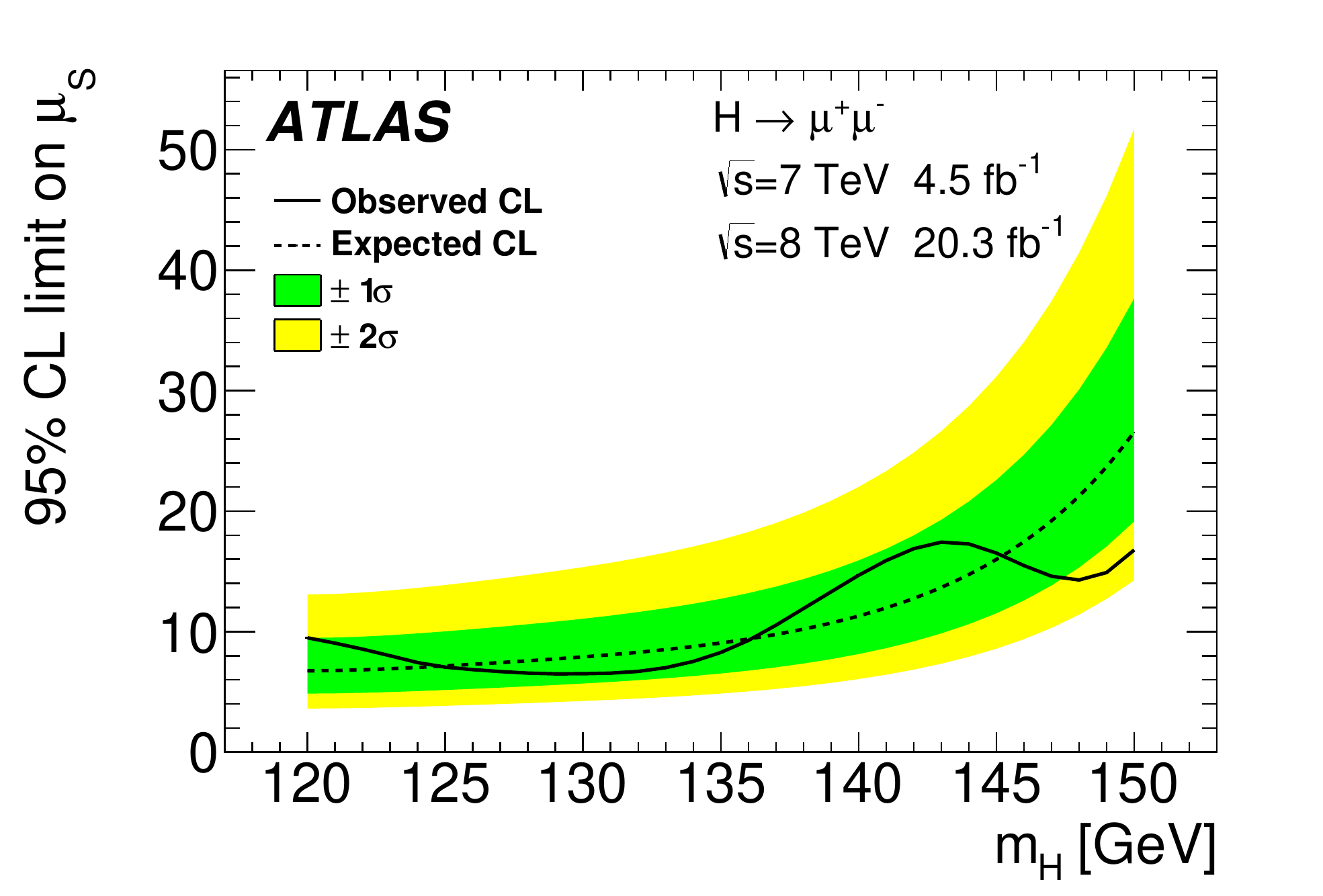}~
  \includegraphics[width=.38\textwidth]{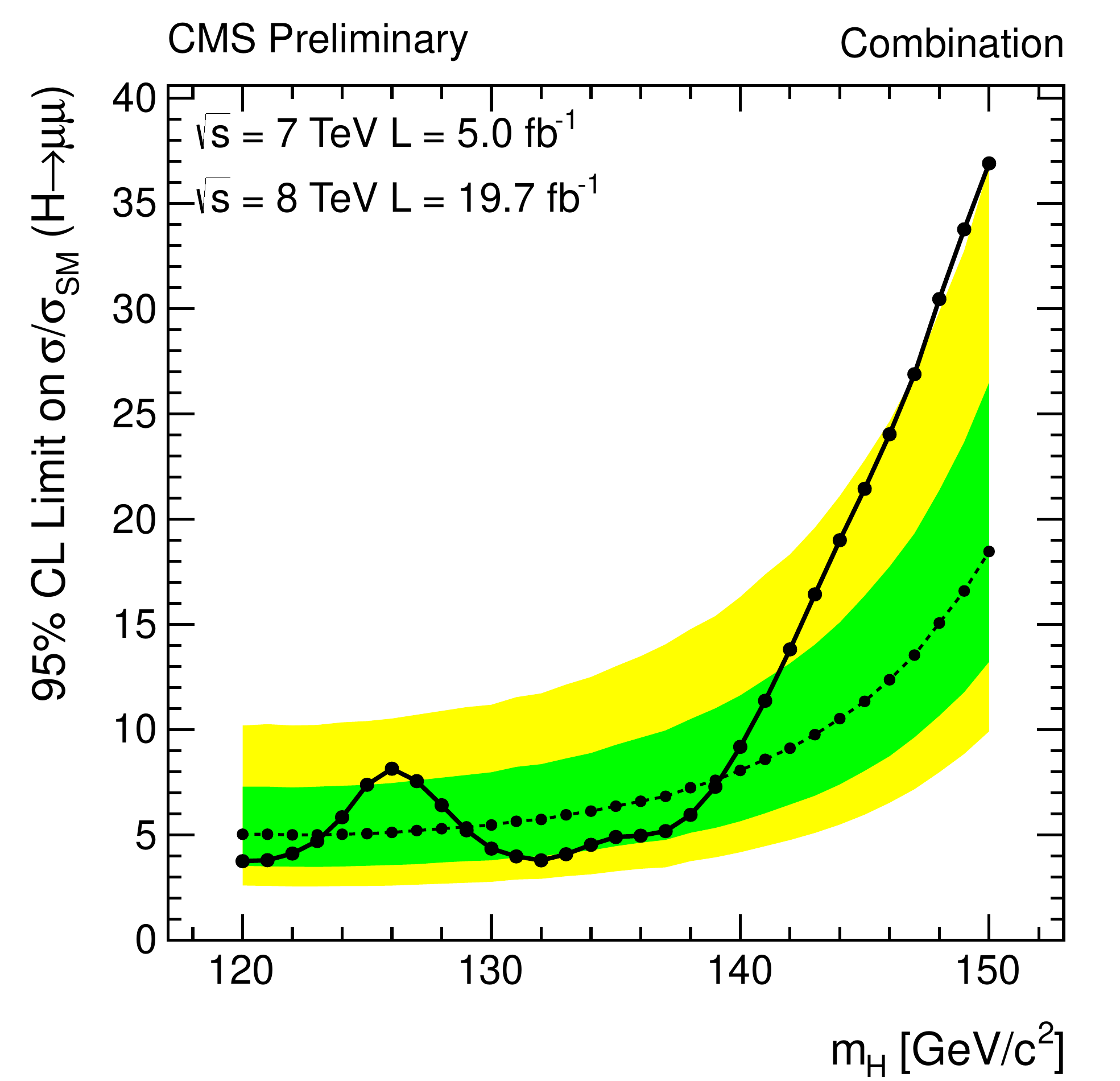}
  \caption{Upper limits on the signal strength of the $H\to\mu\mu$ 
    by ATLAS (left) and CMS (right).}
  \label{fig:Hmumu-lim}
\end{figure}

CMS has a similar analysis and  one can compare the $m_{\mu\mu}$ distribution 
in the equivalent event categories in Fig.~\ref{fig:Hmumu-mass}.
Neither experiment sees any significant excess of events, hence the result is a an upper limit
on the signal strength ($\sim5-8\times$SM for $m_H = 125$\,GeV), see Fig.~\ref{fig:Hmumu-lim}.

\section{ttH}
Search for ttH decay is the hardest among all,
not only because of a tiny signal rate expected from SM, 
but also because of the experimentally difficult final state due to top quarks decay, $t\to Wb$.

\begin{figure}[b]
  \centering
  \includegraphics[width=.45\textwidth]{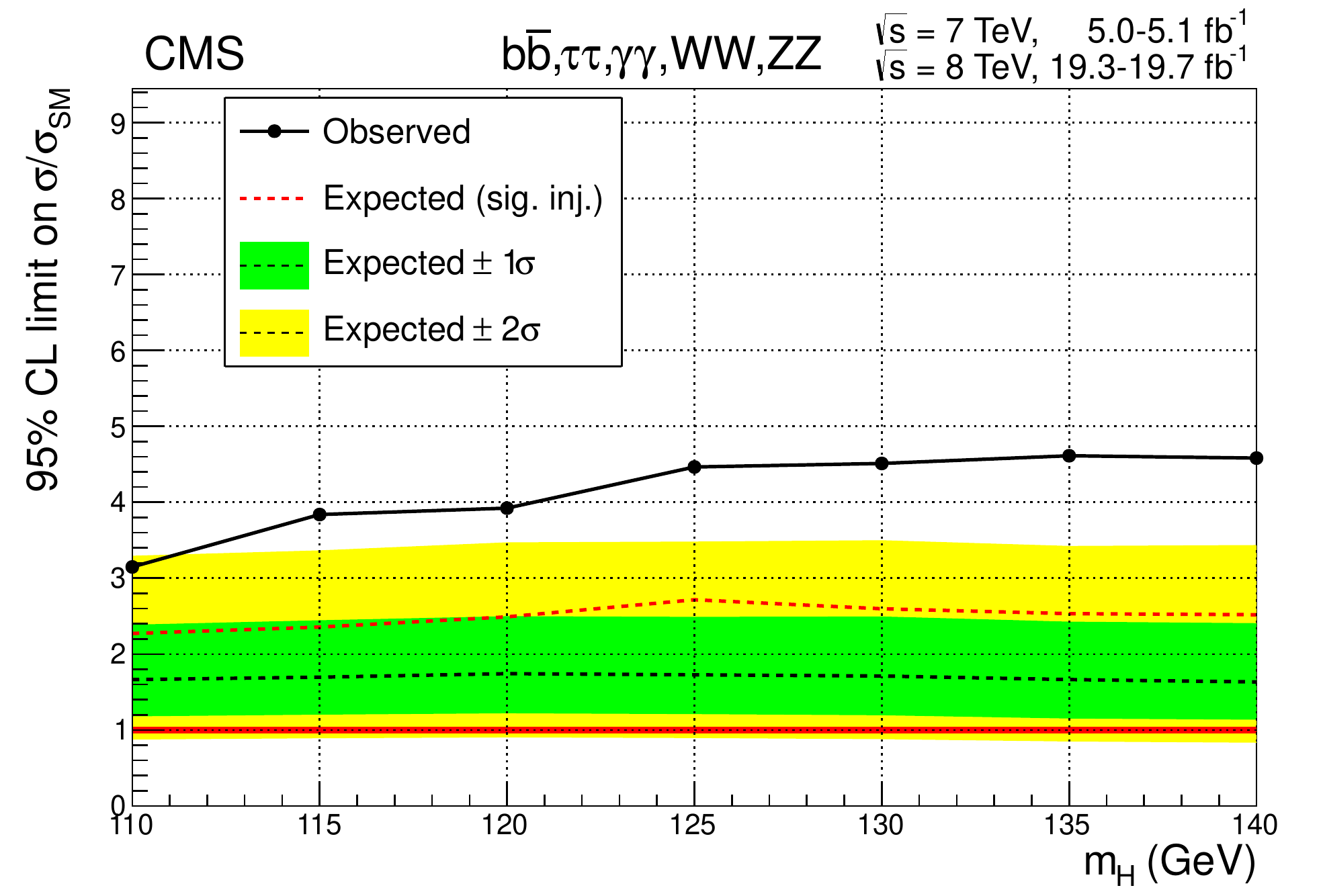}
  \includegraphics[width=.48\textwidth]{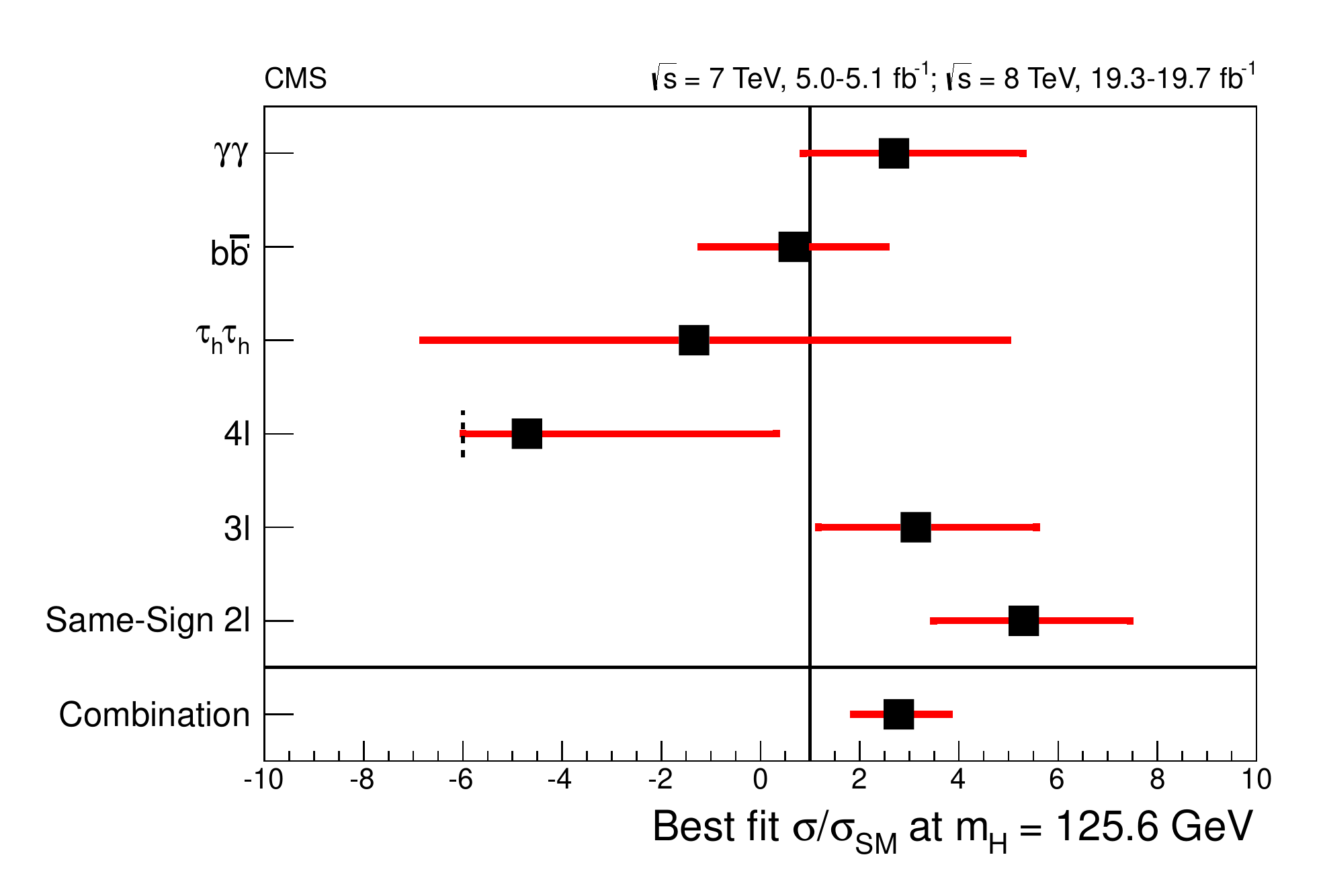}~
  \caption{CMS upper limit on the signal strength of the  ttH  process (left);
    and the best fit $\mu$ by event category (right).}
  \label{fig:ttH-cms}
\end{figure}

In order to see any events from Higgs boson one has to look first for the dominant decay, $H\to b\bar{b}$,
which is the analysis done by ATLAS \cite{atl-ttH} (for $m_H = 125$\,GeV only).
They put an upper limit of $4.1\times$SM with the expected of $2.6\times$SM.
CMS analysis combines $H\to b\bar{b}, \tau\tau, WW, ZZ, \gamma\gamma$ modes.
The upper limit by CMS is set to $4.5\times$SM with the expectation of $1.7\times$SM \cite{cms-ttH}. 
A broad excess across $m_H$ is observed 
and found to be coming from same-sign di-muon events, see Fig.~\ref{fig:ttH-cms}.

\section{Summary}
In spite of the very challenging analyses,
both ATLAS and CMS experiments achieved significant progress in exploring $H\to ff$ decays.
An excess of events with more that $3\sigma$ significance is reported in $H\to\tau\tau$
channel by both experiments -- evidence for $H~\to~\tau\tau$ decay.
The boundaries of $H\to b\bar{b}$ search are pushed
tremendously: expected limit on the signal strength is below one times the SM prediction.

$H\to\mu\mu$ decay and ttH production channel are still waiting for the right time
(or better to say: more statistics) -- we hope to see those peaks in the future!

So far everything is consistent with the Standard Model.


\clearpage

\clearpage
\begin{figure}[t]
  \centering
  \includegraphics[width=.37\textwidth]{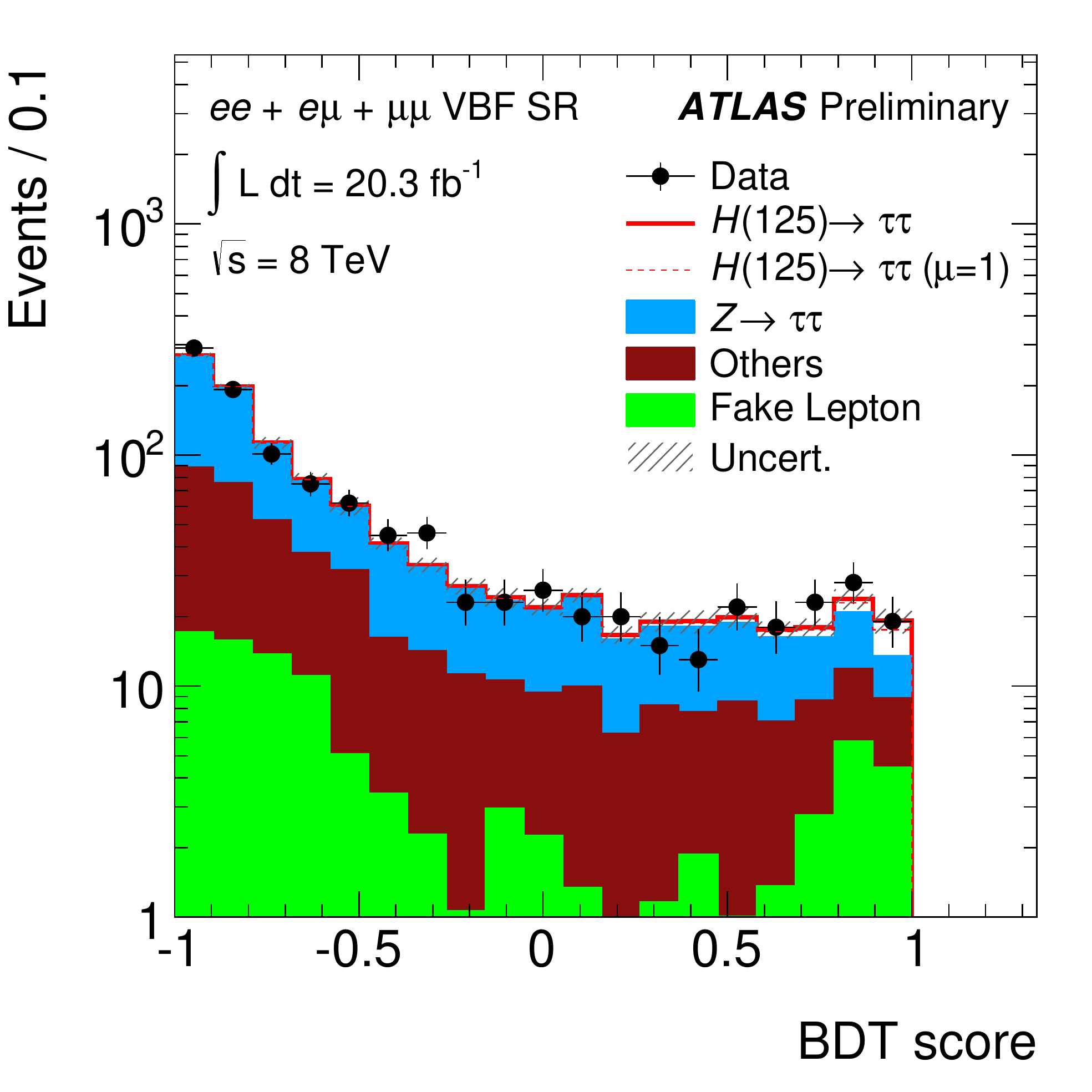}~
  \includegraphics[width=.37\textwidth]{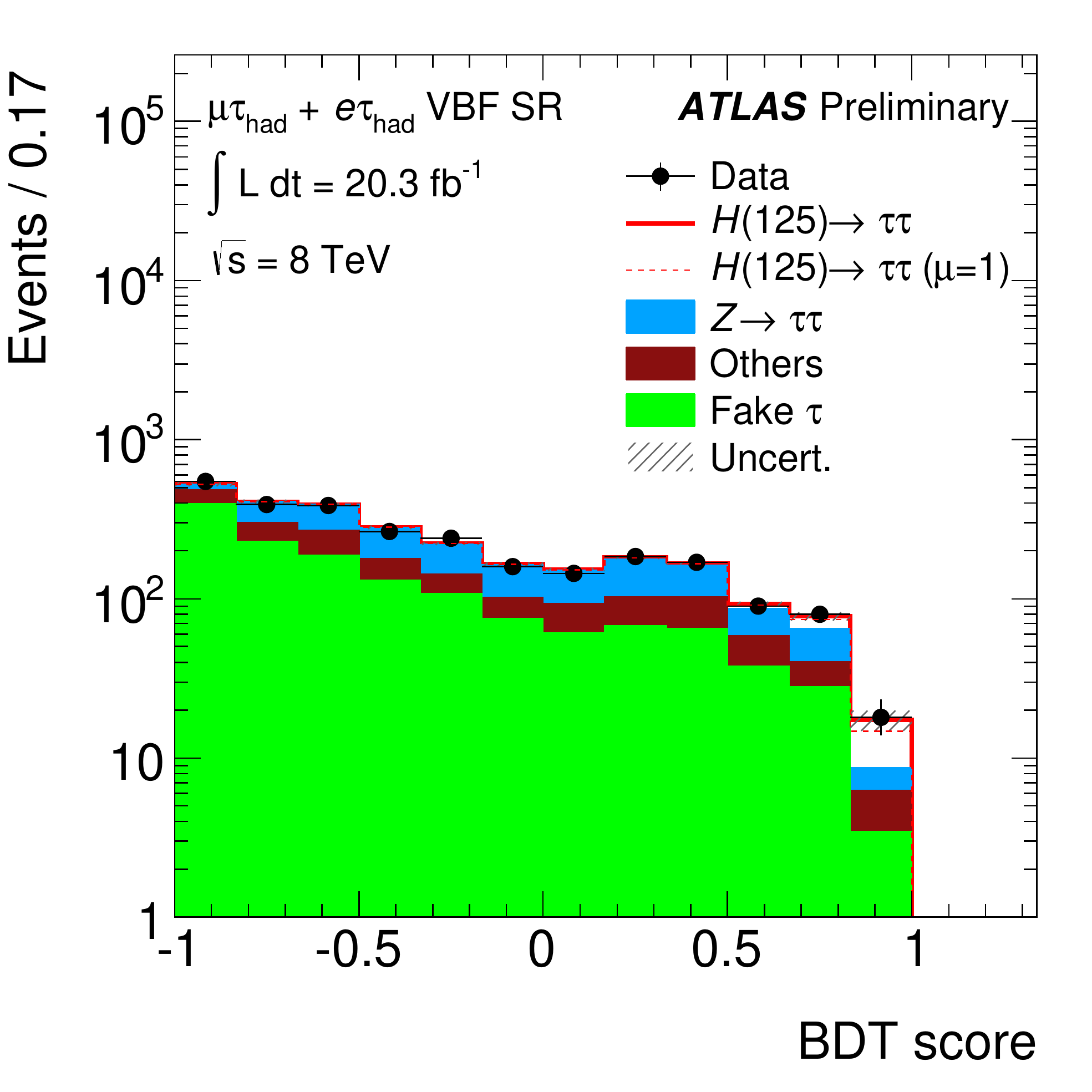}\\
  \includegraphics[width=.37\textwidth]{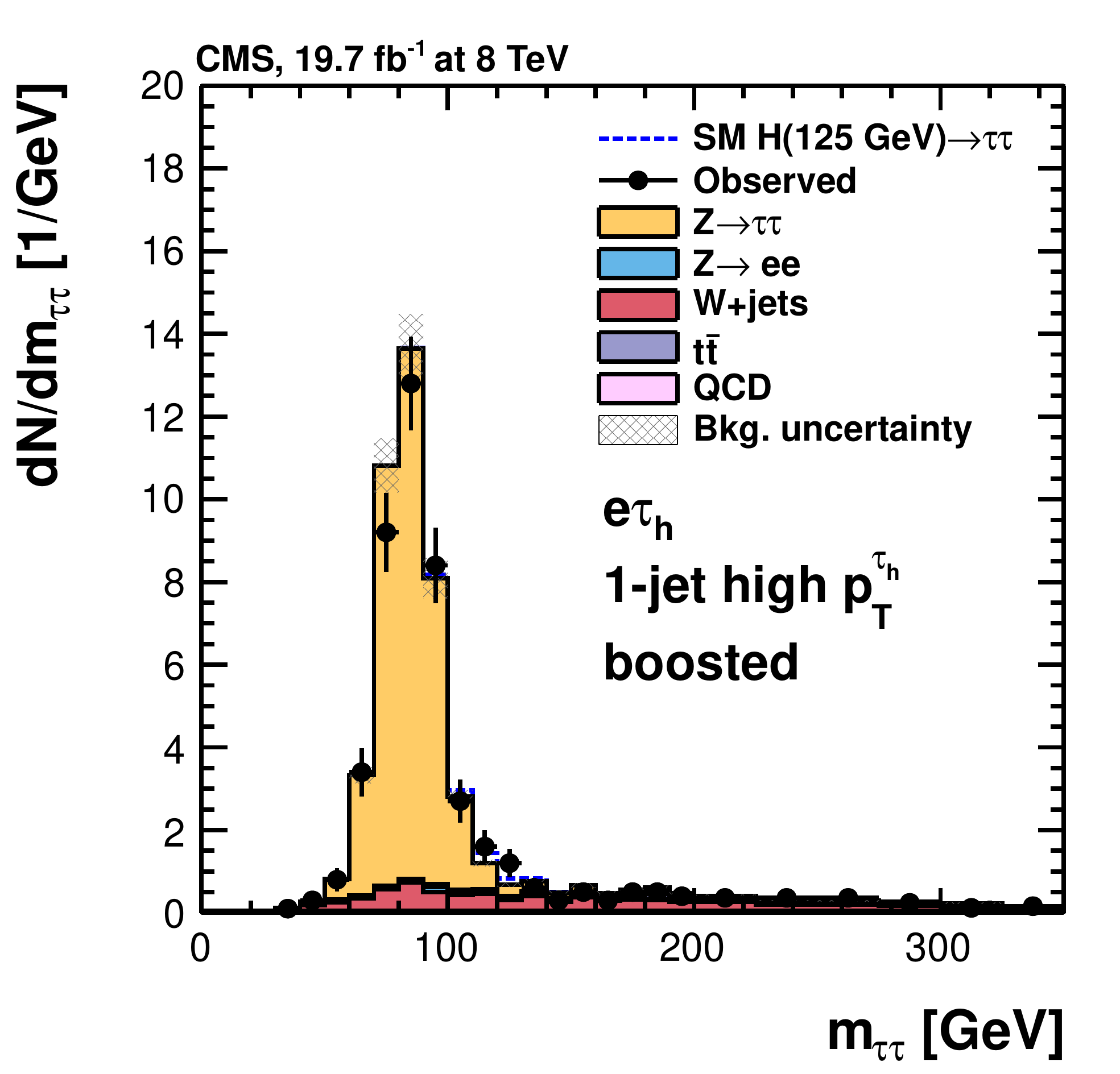}~
  \includegraphics[width=.37\textwidth]{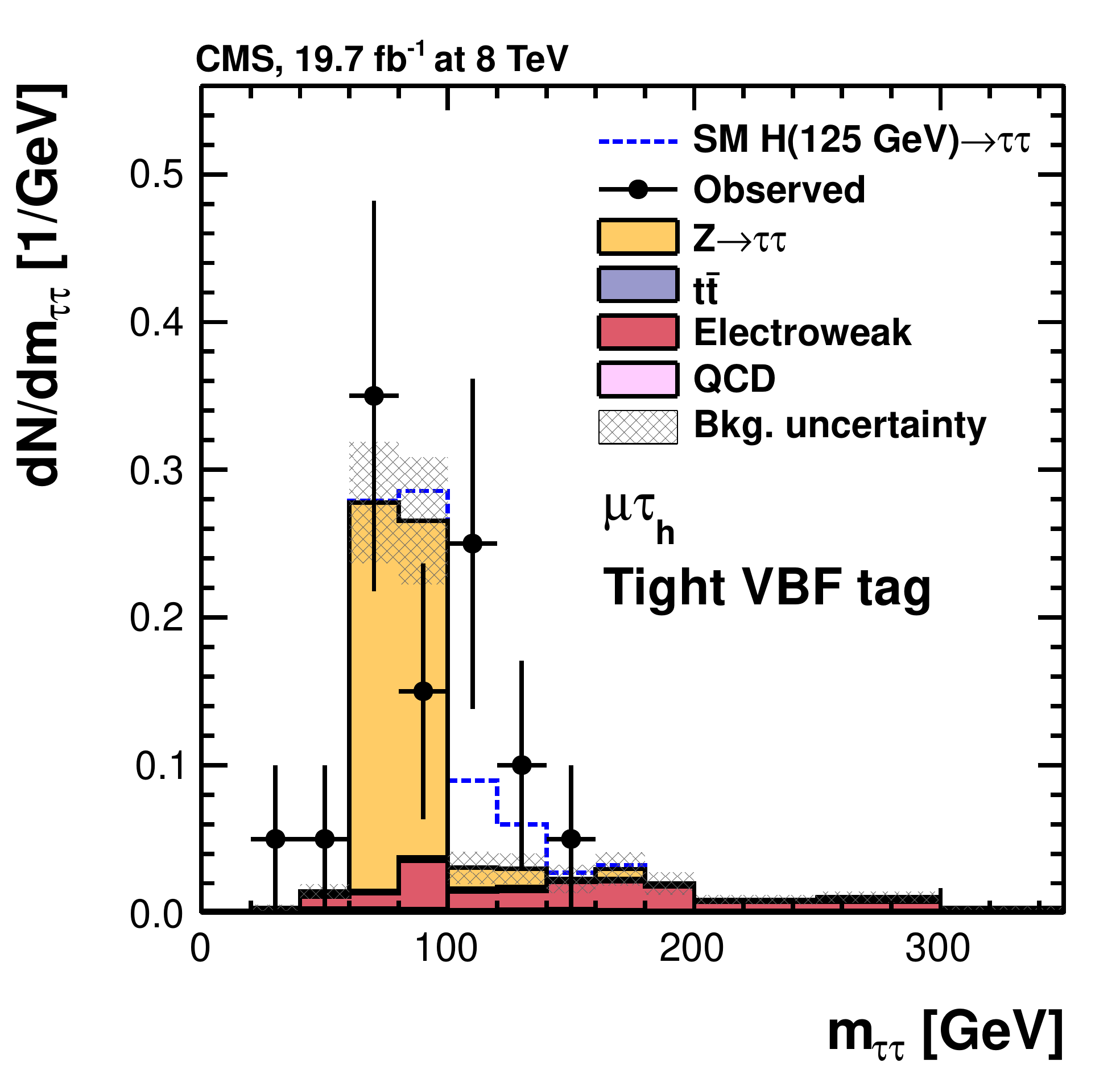}
  \caption{Example distributions of the final  BDT output from ATLAS analysis (top)
  and $m_{\tau\tau}$ from CMS, after the global fit.}
  \label{fig:Htautau-ex}
\end{figure}

\clearpage

\clearpage


\clearpage

\end{document}